\documentclass[twocolumn,showpacs,amssymb]{revtex4}
\usepackage{graphicx}
\usepackage{dcolumn}
\usepackage{bm}

\def\beq{\begin{equation}}
\def\enq{\end{equation}}
\def\ba{\begin{eqnarray}}
\def\ea{\end{eqnarray}}
\def\Meszaros{M\'esz\'aros~}
\def\<{<\!\!}
\def\>{\!\!>}

\begin{document}
\input{epsf}

\title{High-energy Cosmic Rays and Neutrinos from Semi-relativistic
Hypernovae}

\author{Xiang-Yu Wang$^{1,2}$, Soebur Razzaque$^{1,3}$,
Peter \Meszaros$^{1,3}$ and Zi-Gao Dai$^{2}$ }

\affiliation{$^1$Department of Astronomy \& Astrophysics,
Pennsylvania State University, University Park, PA 16802, USA\\ $^2$
Department of Astronomy, Nanjing University, Nanjing 210093, China\\
$^3$Department of Physics, Pennsylvania State University, University
Park, PA 16802, USA}

\begin{abstract}
The origin of the ultrahigh-energy (UHE) cosmic rays (CRs) from
the second knee ($\sim6\times10^{17}$eV) above in the CR spectrum
is still unknown. Recently, there has been growing evidence that a
peculiar type of supernovae, called hypernovae, are associated
with sub-energetic gamma-ray bursts (GRBs), such as
SN1998bw/GRB980425 and SN2003lw/GRB031203. Such hypernovae appear
to have high (up to mildly relativistic) velocity ejecta, which
may be linked to the sub-energetic GRBs. Assuming a continuous
distribution of the kinetic energy of the hypernova ejecta as a
function of its velocity $E_k\propto (\Gamma\beta)^{-\alpha}$ with
$\alpha\sim 2$, we find that 1) the external shock wave produced
by the high velocity ejecta of a hypernova can accelerate protons
up to energies as high as $10^{19}~{\rm eV}$; 2) the cosmological
hypernova rate is sufficient to account for the energy flux above
the second knee; and 3) the steeper spectrum of CRs at these
energies can arise in these sources. In addition, hypernovae would
also give rise to a faint diffuse UHE neutrino flux, due to
$p\gamma$ interactions of the UHE CRs with hypernova optical-UV
photons.

\end{abstract}

\maketitle

There is a general consensus that galactic supernova remnants
(SNRs) are responsible for the CRs at energies below the ``knee''
at $\sim 3\times 10^{15}$~eV \cite{galactic-origin}, most probably
through the diffusive shock acceleration mechanism
{\cite{diffuse-shock}}.  Galactic SNRs expanding into their former
stellar wind have been suggested to be responsible for CRs above
the knee {\cite{galactic-wind}}. Recent data from the KASCADE
experiment suggest that heavy elements of nuclear charge $Ze$ are
accelerated by the galactic SNRs  to the magnetic rigidity limit
$\sim 3\times 10^{15}Z$~{eV} \cite{kascade}. Thus galactic SNRs
may be able to produce CRs to at least $\sim 10^{17}$~eV (see e.g.
\cite{NW00, Hillas05} for recent reviews), an energy slightly
below the ``second knee'' in the CR spectrum at $\sim
6\times10^{17}$~eV (see e.g. \cite{2nd-knee}). On the other hand,
the highest energy CRs above a few times $10^{19}~{\rm eV}$ are
generally thought to be extra-galactic in origin, due to their
isotropic distribution and a lack of galactic source candidates
capable of producing them. At these energies, the possible sources
include cosmological GRBs \cite{CR-GRB}, active galactic nuclei
\cite{Berezinsky06} and  powerful radio galaxies
\cite{radio-galaxies}.

The origin of the intermediate energy range CRs,
$10^{17}-10^{19}$~eV, however, remains more elusive. Some authors
have suggested a common galactic origin for all CRs between the
first knee and $10^{20}$~eV in young neutron stars or Magnetars
\cite{gal-CR}, while others favor an extra-galactic origin of all
CRs above $10^{18}$~eV \cite{Berezinsky06, Hillas05, Gaisser}.
Recent HiRes data shows that the transition from heavy nuclei to
the proton composition may already occur at the second knee
\cite{composition}, suggesting that the UHE  cosmic-rays with
energy from $6\times10^{17}$eV  above are all probably
extra-galactic \cite{Hillas05, Gaisser}. In this paper we show
that the energetics and spectrum of CRs from the second knee to
$10^{19} $eV may be due to extra-galactic hypernovae, similar to
the hyper-energetic supernova SN1998bw.

SN1998bw was striking not only in its unusually large explosion
energy, $E\simeq 3-5\times10^{52}{\rm erg}$ {(so called
``hypernovae" \cite{Paczynski98})}, but also in that it was
associated with a very sub-energetic GRB, GRB980425, with an
isotropic equivalent gamma-ray energy $E_\gamma\simeq 10^{48}~{\rm
erg}$ \cite{Galama98}. The possible transition from extreme SNe to
GRBs was implied in the magneto-rotational mechanism of the SN
explosions proposed in Ref.\cite{Bisnovatyi-Kogan70,
Bisnovatyi-Kogan0506}.  Now the connection between GRB and SNe has
been established through observations. {The hypothesis for
SN-associated GRBs such as GRB980425 that their weakness is an
apparent effect due to seeing the explosion off-axis runs counter
to observational tests based on long-time radio observations
\cite{Soderberg04}, so it is likely that this GRB is inherently
dimmer than typical. } The observations of the radio afterglow of
this event showed that about $10^{50}~{\rm erg}$ of kinetic energy
were released in the form of a mildly relativistic ejecta
{\cite{98bw-radio}}. Due to the large supernova explosion energy
and the much lower than typical GRB energy, attempts have been
made to ascribe the GRB event to the shock from the mildly
relativistic ejecta as it breaks out through the hypernova
progenitor's outer envelope {\cite{Woosley99, Tan01}}. A generally
accepted conclusion, however, has not yet been reached. A recently
detected strong thermal X-ray emission component in another
sub-energetic burst (GRB060218), associated with SN2006aj, may
also be associated with a semi-relativistic supernova shock
breakout, in which the mildly relativistic supernova ejecta has an
energy $\gtrsim10^{49}~{\rm erg}$ {\cite{Campana}}. The radio
observations of this burst, as well as those of another hypernova
burst, GRB031203/SN2003lw, also indicate that there is a
significant energy in the mildly relativistic ejecta
{\cite{SN2003lw}}.  We will use the term semi-relativistic
hypernovae to denote such supernovae exhibiting a mildly
relativistic ejecta component, seen in association with GRBs. {The
recently discovered SN2006gy \cite{2006gy} also has a large
explosion energy, but continued multi-wavelength monitoring has
not yielded any evidence for an associated GRB, so this object may
be in the class of normal hypernovae with large explosion energy,
but without any significant mildly-relativistic ejecta
\cite{nomoto06}. }
\begin{figure}
\centerline {\epsfxsize=3.5in \epsfbox{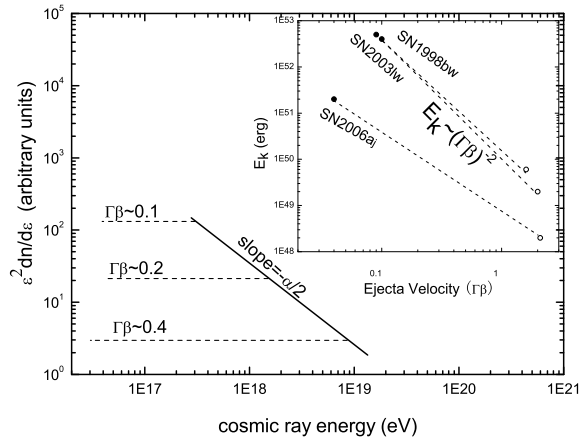}} \caption{The
expected spectrum of CRs as a function of energy $\varepsilon$
produced by hypernova remnants with a distribution of the ejecta
kinetic energy with velocity $E_k\propto (\Gamma\beta)^{-\alpha}$.
Dashed lines indicate the flat, injection spectrum
($\varepsilon^2dn/d\varepsilon \propto \varepsilon^0$) from single
velocity ejecta for three different velocities.  The convolved
contribution from different velocity ejecta, denoted by the solid
line, leads to a spectrum $\varepsilon^2dn/d\varepsilon\propto
\varepsilon^{-\alpha/2}$. The inset shows the kinetic energy
distribution of three nearby hypernovae associated with
sub-energetic GRBs. The data points are from
Ref.{\cite{Soderberg06}}. The solid and blank circles denote the
energy of the slowest ejecta and the mildly relativistic ejecta,
respectively. A kinetic energy distribution $E_k\propto
(\Gamma\beta)^{-\alpha}$ (the solid line) gives a rough fit to the
data of SN1998bw/GRB9802425 and SN2003lw/GRB031203 with
$\alpha\sim2.4$, while for SN2006aj/GRB060218, the slope is
$\alpha\sim-1.7$.} \label{fig:distribution}
\end{figure}

In Figure 1 (inset) we show the kinetic energy distribution of the
supernova ejecta associated with these three nearby sub-energetic
GRBs, ranging from the bulk of the ejecta at $\Gamma\beta\simeq
0.1 $ to the mildly relativistic ejecta ($\Gamma\beta\simeq1$),
where $\beta=v/c$ and $\Gamma$ are the ejecta normalized velocity
and bulk Lorentz factor, respectively. Surprisingly, even though
the energy estimates of the high velocity ejecta from the radio
observations are crude, all three hypernovae give a roughly
consistent extrapolation of the slope of the kinetic energy
distribution from the low to the high velocity end.  SN1998bw and
SN2003lw give a consistent slope of about $\sim-2.4$, while
SN2006aj gives a slightly shallower distribution with a slope of
$\sim -1.7$.  Note that if the explosion is aspherical, the
kinetic energy released in SN1998bw may be lower, $\sim
2\times10^{52}{\rm erg}$ \cite{98bw-energy}, which will lead to a
slightly shallower slope. It has been shown in \cite{Soderberg06}
that the relatively shallow decay of the radio afterglow of
GRB060218/SN2006aj can be modelled with a shock expansion
$r\propto t^{0.85}$, appropriate for a core-collapse supernova
explosion with a continuous distribution of ejecta velocities
{\cite{Chevalier98}}, propagating into a stellar wind environment
of density $\rho\propto r^{-2}$. This provides a plausible
scenario for a continuous distribution of the ejecta kinetic
energy over velocities, ranging from the low velocity (0.1c)
supernova bulk ejecta to the mildly relativistic ejecta within the
same explosion, of the form
\begin{equation}
E_k\propto (\Gamma\beta)^{-\alpha}.
\end{equation}
{Such a distribution of velocities is naturally expected in an
outflow spreading over a wide range of angles
\cite{2006aj-theory}. }

Standard hydrodynamic collapse calculations involving
non-relativistic shocks result in a kinetic energy profile
$E_k\propto (\Gamma\beta)^{-5}$ {\cite{Tan01}}, with a negligible
fraction of the kinetic energy at mildly relativistic velocities,
consistent with the radio observations of, e.g., local type Ib/c
supernovae 1994I and 2002ap. This very steep velocity profile
implies negligible contribution to the highest energy CRs by high
velocity ejecta \cite{bv04}. On the other hand, ultra-relativistic
shocks result in a much flatter profile, $E_k\propto
(\Gamma\beta)^{-\alpha}$ with $\alpha\simeq1$ {\cite{Tan01}. For
shocks in the trans-relativistic velocity regime, the energy
distribution has not been calculated, but the above assumed slope
of $\alpha\sim2$ seems to be intermediate between the two extreme
regimes. An important implication of such a continuous energy
distribution in the semi-relativistic regime is that there is a
significant amount of energy in the high-velocity ejecta of a
hypernova. At this high velocity, the hypernova blast wave could
accelerate CRs to energies as high as $10^{19}$~eV, as we show
below.

{\em Maximum energy of accelerated CRs. ---} Diffusive shock
acceleration in supernova shock fronts has been extensively
studied. The maximum energy for CR acceleration by the SNRs is
usually thought to be limited to $\sim 10^{14}-10^{15}$~eV for the
case of typical interstellar magnetic fields of a few $\mu {\rm
G}$. {It was suggested in Refs.\cite{galactic-wind} that the
stellar wind from Wolf-Rayet (WR) star  may have a relatively high
magnetic field with a dominant component transverse to the shock
normal, so that the SN explosion in these winds can accelerate
particles to a much larger maximum  energy.} Recently there are
also suggestions  (e.g. \cite{BL01}) that  the magnetic field can
be amplified non-linearly through MHD turbulence excited by the
CRs in the vicinity of the shock to many times the pre-shock
values, thus significantly increasing the acceleration rate and
hence increasing the maximum energy. Amplification of the magnetic
field to an equipartition value is generally assumed in radio SNRs
and in GRB afterglow shocks, and has gained support from recent
X-ray observations of several young SNRs\cite{B-snr}. We here
consider a semi-relativistic hypernova ejecta with a velocity
distribution given by Eq.(1), expanding in the stellar wind
characteristic of WR stars, which are  thought to be the
progenitors of the hypernovae associated with GRBs. {Different
from Refs.\cite{galactic-wind}, we here consider a random,
CR-amplified magnetic field with a strength close to the
equipartition value.} During the free expansion phase, the
magnetic field is $B^2/8\pi=2\epsilon_B \rho_w(R) c^2 \beta^2$,
where $\epsilon_B=0.1\epsilon_{B,-1}$ is the fraction of the
equipartition value of the magnetic field energy and $\rho_w$ is
the mass density of the stellar wind at radius $R$. The magnetic
field at the free-expansion radius $R$ is
\begin{equation}
\begin{array}{ll}
B=0.03\epsilon_{B,-1}^{1/2}R_{18}^{-1}\left(\frac{v}{10^{10}{\rm
cm s^{-1}}}\right)\left(\frac{\dot{M}}{3\times10^{-5} {\rm M_\odot
yr^{-1}}}\right)^{1/2}\\
\times v_{w,3}^{-1/2} {\rm G},
\end{array}
\end{equation}
where ${\dot{M}}$ is the wind mass loss rate, whose average value
is $3\times10^{-5} {\rm M_\odot yr^{-1}}$ for WR stars, and $v_w =
10^3 v_{w,3}$~km/s is the wind velocity. {The maximum energy that
can be attained in a shock depends on the magnetic field
configuration. If the magnetic field is mostly perpendicular to
the shock surface, the scattering coefficient for particles is
smaller than in the case of a parallel shock and as a result, the
maximum energy is given by $\simeq ZeBR$ \cite{Jokipii87}. In our
case, however, we assume that the CR-amplified magnetic field is
random in direction and }  the maximum energy of the accelerated
particles is (e.g. \cite{Hillas84, BL01})
\begin{equation}
\begin{array}{ll}
\varepsilon_{\rm max}\simeq Z e BR\beta = 4\times10^{18} Z \\
\times\epsilon_{B,-1}^{1/2}\left(\frac{v}{10^{10}{\rm cm
s^{-1}}}\right)^{2}\left(\frac{\dot{M}}{3\times10^{-5} {\rm
M_\odot yr^{-1}}}\right)^{1/2}v_{w,3}^{-1/2} {\rm eV}.
\end{array}
\end{equation}
The stellar wind of WR stars is composed largely of H and He
($Z=1$ and $Z=2$, respectively). In this shock acceleration
scenario, the maximum energy $\varepsilon_{\rm max}$ is
proportional to the square of the shock velocity, so a higher
velocity hypernova ejecta can lead to a higher $\varepsilon_{\rm
max}$. Note also that during the free expansion phase of the
ejecta, $\varepsilon_{\rm max}$ is independent of the radius. For
the assumed velocity distribution of Eq.(1), the bulk of the
ejecta has a velocity of $0.1c$ and the maximum CR energy
corresponding to this (low-end) velocity ejecta is about
$10^{17.5}Z~ {\rm eV}$ for typical parameters of the stellar wind.

{\em The spectrum of the CRs.  ---} {For a single velocity ejecta,
the differential spectrum of the accelerated protons is given by
the injection spectrum, which is  $dN/d\varepsilon\propto
\varepsilon^{-\gamma}$ with $\gamma\simeq 2.0$, for both
non-relativistic shocks and semi-relativistic shocks
\cite{spectrum-index}.} However, if the hypernova produces a
kinetic energy distribution spread over the velocity with the same
explosion, as described by Eq.(1), the final CR spectrum detected
at Earth is determined by the kinetic energy distribution profile,
as illustrated in Fig.1. {This can be understood as higher energy
CRs being contributed dominantly by higher velocity ejecta, which
are represented with  a smaller total energy.}
As the maximum CR energy for a particular velocity is
$\varepsilon_{\rm max}\propto (\Gamma\beta)^2$ and the energy
distribution of the ejecta is $E_k\propto
(\Gamma\beta)^{-\alpha}$, we see that $E_k\propto \varepsilon_{\rm
max}^{-\alpha/2}$. Convolving the contribution from the different
velocity ejecta (Fig.1), we expect a final differential energy
spectrum of the CRs of the form
\begin{equation}
\varepsilon^{2}(dN/d\varepsilon)\propto \varepsilon^{-\alpha/2},
\label{eq:convolved-flux}
\end{equation}
{where $\varepsilon$ is the energy of cosmic ray particles, which
relates to the ejecta kinetic energy as $E_k(\varepsilon)\propto
\varepsilon^{2}(dN/d\varepsilon)$.} Fits to the observed CR data
give a differential spectrum $J=C(E/6.3\times10^{18}{\rm
eV})^{-3.20\pm0.05}$ for $4\times10^{17}{\rm
eV}<E<6.3\times10^{18}{\rm eV}$ and $J=C(E/6.3\times10^{18}{\rm
eV})^{-2.75\pm0.2}$ for $6.3\times10^{18}{\rm
eV}<E<4\times10^{19}{\rm eV}$ with $C=(9.23\pm0.65)\times10^{-33}
{\rm m^{-2} s^{-1} sr^{-1} eV^{-1}}$ \cite{NW00}. The observed
spectral slope at energies $4\times10^{17}{\rm eV}<E<
6.3\times10^{18}{\rm eV}$ implies $\alpha=2.40\pm0.1$ in Eq.
(\ref{eq:convolved-flux}), which is roughly consistent with the
theoretical slope deduced in Fig.1. Note that the energy losses
due to photo-pair production with CMB may become important above
$\sim10^{18}$~eV and thus steepening the injection spectrum
\cite{Berezinsky06}.  However, the details depend on the specific
value of the injection spectrum and source evolution.  In all
cases this is a small correction which  can be accommodated  by
using a slightly smaller value for $\alpha$ and thus a slightly
harder injection spectrum from the HNe.

{\em Hypernova rates and the observed flux ---} We estimate now
how many SN1998bw-like hypernovae per unit volume per unit time
are needed to produce the CR flux from $6\times10^{17}$ to
$10^{19}$~eV. Assuming that the kinetic energy output from one
SN1998bw-like hypernova is $E_{k,{\rm HN}}=5\times10^{52}~{\rm
erg}$ , the local kinetic energy release rate by hypernovae is
\begin{equation}
\begin{array}{ll}
\dot{\epsilon_k}(z=0)=R_{\rm HN}E_{k,{\rm HN}} \\
=2.5\times10^{46} \left(\frac{R_{\rm HN}}{500 {\rm Gpc^{-3}
yr^{-1}}}\right) {\rm erg~ Mpc^{-3}~ yr^{-1}}
\end{array}
\end{equation}
Adopting an efficiency factor $1/6$ for the conversion of ejecta
kinetic energy into CR energy \cite{Hillas05}, and 1/${\rm ln}
(\varepsilon_{max}/\varepsilon_{min})\simeq0.1$ as the fraction of
the total CR energy that is contributed by each decade of energy,
the local CR energy generation rate per energy decade at
$10^{17.5}Z$~eV, corresponding to $v=0.1c$, is
$\dot{\epsilon}_{CR,0}=0.016\dot{\epsilon_k}(z=0)$. The
corresponding expected CR flux is
\begin{equation}
 \varepsilon^2J=({c}/{4\pi H_0})\dot{\epsilon}_{CR,0}
 f_z
\end{equation}
where $H_0=70$~km~s$^{-1}$~Mpc$^{-1}$ is the Hubble constant and
\begin{equation}
f_z=H_0\int_0^{z_{\rm max}} dz ({dt}/{dz}) S(z) (1+z)^{-1}
\end{equation}
is the correction factor for the contribution from high-redshift
sources. {Here $({dt}/{dz})=H^{-1}(z)/(1+z)$, with $H(z)$ being
the Hubble parameter at cosmological epoch $z$, and $z_{\rm max}$
is the maximum redshift corresponding to the mean free path
against photopion production in the CMB, whose value is $z_{\rm
max}\gtrsim4$ for protons with energies $\lesssim 10^{19}$eV.}
 The
value of $f_z$ is $\sim 2-3$ for $z_{\rm max} = 4$ and for the
source evolution function $S(z)$ given by different star formation
rates (SFR) in Ref.~\cite{sfr} or the broken power-law estimate in
Ref.~\cite{wblimit} for a standard $\Lambda$CDM cosmology.  Here
we assume $f_z \approx 3$. At $\varepsilon=10^{17.5} Z$~eV, we get
a CR flux of
\begin{equation}
J=10^{-28}Z^{-2}\left(\frac{R_{\rm HN}}{500 {\rm Gpc^{-3}
yr^{-1}}}\right) \left(\frac{f_z}{3}\right) ~{\rm eV^{-1} m^{-2}
s^{-1} sr^{-1}}
\end{equation}
Comparing this to the observed CR flux of $1.5\times10^{-28}
(\varepsilon/10^{17.5}~{\rm eV})^{-3.2}$, we infer a required
hypernova rate of
\begin{equation}
R_{\rm HN}=750 Z^{-1.2}({f_z}/{3})^{-1}{\rm Gpc^{-3} yr^{-1}}
\end{equation}

Assuming Z=1 (or 2) and $f_z=3$, one can derive a required
hypernova rate of $330-750$ ${\rm Gpc^{-3} yr^{-1}}$. Comparing
this with the local rate of ``normal" type Ib/c SNe, $\sim
2-5\times10^{4}~{\rm Gpc^{-3} yr^{-1}}$ {\cite{Guetta07, Ib/c},
one can find that the ratio of the required hypernovae rate
 to the normal Ib/c SNe rate is  $\sim1-4 \%$, which is
consistent with the value observed in the local universe $\sim
7\%$ \cite{Guetta07}}. The required semi-relativistic hypernova
rate is also consistent with the observed rate of low-luminosity
GRBs \cite{Soderberg06, Liang06, Guetta07}. Since different SFR
give different values of $f_z$, one can in principle use  the
required hypernova rate to constrain $f_z$ and therefore constrain
the SFR.

{\em Neutrino emission from CRs interacting with hypernova
photons.---} The kinetic energy of the highest velocity ejecta of
hypernovae is converted into the highest energy CRs in a
relatively short time, given by the free expansion time $t<t_{\rm
dec}$ before the ejecta is decelerated by the swept-up stellar
wind. During this time the hypernovae remain very bright in the
optical-UV band. As a consequence, the high-energy protons can
interact with the hypernova optical-UV photons during the first
tens of days after the explosion, leading to $p\gamma$ neutrino
production. This neutrino production mechanism differs in its
origins from another, recently suggested neutrino production
scheme in low-luminosity GRBs \cite{neutrino-ll}, which involves
the usual internal shock model for the acceleration of protons and
production of the target photons. {It also differs from the
suggested mechanism of neutrino production in starburst galaxies,
where neutrinos are thought to arise from $pp(pn)$ interactions
between the accelerated protons and the interstellar medium
\cite{starburst-neutrinos}. }

Around the peak photon luminosity time (typically a few tens of
days) of type Ib/c supernovae, only the high velocity ejecta with
$\Gamma\beta\gtrsim1$ has been decelerated, and only about
$2\times 10^{50}~{\rm erg}$ of energy goes into such high velocity
ejecta. The resulting high-energy protons are above the photopion
interaction threshold with thermal hypernova photons if their
energy satisfies $\varepsilon_p \varepsilon_\gamma \gtrsim 0.3
{\rm GeV^2}$, i.e. $\varepsilon_p\gtrsim10^{18}{\rm eV}$ for
UV-optical photons. The cooling rate of a proton of energy
$\varepsilon_p$ due to pion production via $\Delta^+$ resonance is
$t_{p\gamma}^{-1}\equiv E^{-1}(dE/dt) \simeq
n(\varepsilon_\gamma)\sigma_{peak}c\xi_{\rm peak}
(\Delta\varepsilon/\varepsilon_{peak})$.  Here
$n(\varepsilon_\gamma)$ is the number density of hypernova photons
at the peak of the black-body distribution,
$\sigma_{peak}=5\times10^{-28} ~{\rm cm^{-2}}$ is the cross
section for pion production for a photon with energy
$\epsilon=\varepsilon_{peak} = 0.3~{\rm GeV}$ in the proton rest
frame, $\xi_{peak}=0.2$ is the average fraction of energy lost to
the pion, and $\Delta\varepsilon=0.2{\rm GeV}$ is the peak width.
Thus, the fraction of energy lost by protons to pions is
\begin{equation}
\begin{array}{ll}
 f_{p\gamma}=R/(\Gamma\beta c
t_{p\gamma}) \\
=0.2 L_{\rm SN,43}(R/10^{16}~{\rm
cm})^{-1}(\varepsilon_\gamma/1~{\rm eV})^{-1}
\end{array}
\end{equation}
where $L_{\rm SN}\simeq10^{43}~{\rm erg ~s^{-1}}$ is the bolometric
luminosity of SN1998bw around its peak brightness (see e.g.
{\cite{Galama98}}) and $\varepsilon_\gamma=1$~eV is taken as the
optical photon energy. The $\nu_\mu$ ($\bar\nu_\mu$) from pion decays
will have a typical energy $0.05\varepsilon_p$ and the expected diffuse
neutrino flux is
\begin{equation}
\begin{array}{ll}
\varepsilon_\nu^2 \Phi_{\nu_\mu}\simeq \frac{1}{8} f_{p\gamma}
(\varepsilon_p^2J_p)_{(\Gamma\beta\ge1)}  =0.3\times10^{-10} \\
\times\left(\frac{R_{HE}}{500 {\rm Gpc^{-3} yr^{-1}}}\right)
\left(\frac{f_{p\gamma}}{0.2}\right)\left(\frac{f_z}{3}\right){\rm
~GeV ~cm^{-2} s^{-1} sr^{-1}}
\end{array}
\end{equation}
For the thermal spectrum of the target photons, only a small
energy range of protons with $\varepsilon_p>10^{18}{\rm eV}$ can
interact effectively with the photons, so we expect the neutrino
diffuse emission to peak at $\sim 5\times10^{16}~{\rm eV}$. The
maximum neutrino energy would be $1.5\times 10^{18}$~eV
corresponding to the maximum CR energy of $3\times 10^{19}$~eV.
The probability for a high-energy muon neutrino to interact in
ice/water producing a high-energy muon detected by the embedded
instruments is $P(\varepsilon_\nu) \approx 4.6\times10^{-4}
(\varepsilon_\nu/{\rm PeV})^{0.55}$ in the PeV-EeV range
\cite{nu-detect}. Integrating the flux in Eq.(11) times the
probability in the $5\times 10^{17}-1.5\times 10^{18}$~eV range,
we get an event rate of $\approx 10^{-2} ~{\rm yr}^{-1} ~{\rm
km}^{-2}$ over $2\pi$~sr in a neutrino telescope. Including muon
anti-neutrinos would increase the rate by a factor $2$. This rate
is too low for cubic kilometer detectors, but future $\gtrsim
100~{\rm km}^2$ telescopes such as ANITA and ARIANNA may detect
these ultra-high energy neutrinos \cite{detector}.

 {\em Discussion.---} We have proposed that cosmic rays from the second knee to
$10^{19}$ eV, whose origin has been debated, may be produced by
extra-galactic hypernovae similar to SN1998bw, which are
associated with under-energetic GRBs.
The CRs below the second knee  may be due to  heavy ions
accelerated in Galactic sources, such as  Galactic supernovae
expanding into the ISM \cite{Hillas05}} or stellar winds
\cite{galactic-wind}, and Galactic trans-relativistic hypernovae
\cite{Waxman07}. This is supported by the  measurements from
KASCADE \cite{kascade} that above the ``knee" at $3\times10^{15}$
eV the chemical composition is increasingly richer in heavy
nuclei. Recent HiRes measurements \cite{composition}  found that
the chemical composition  changes back towards  lighter (proton)
composition at and above the "second knee", suggesting a
transition from Galactic CRs to extra-galactic CRs that consist
primarily of protons.  A smooth transition in the CR spectrum
between Galactic and extra-galactic components may be reasonable,
according to the numerical estimates by Hillas (see Fig.8 of
\cite{Hillas06}), who found that the rapidly falling spectrum may
accommodate a factor of 3 or more uncertainty in the
extra-galactic component without providing a visible clue to the
joint point of the overall spectrum. {Berezinsky et
al.\cite{Berezinsky06} also considered an extra-galactic origin of
the cosmic-rays above the second knee, similar to us, but with a
single component extending up to the highest energy that
originates from AGNs.  The so-called "dip" between the second knee
and the ankle, i.e. first steepening and then hardening of the
spectrum, has been suggested as being due to adiabatic and
pair-production energy losses by cosmic-ray protons. In a
multi-component model such as ours, the "dip" can arise due to a
steep injection spectrum, from HNe above the second knee (modulo
small adiabatic and pair-production energy losses) which is taken
over by a harder spectrum from normal GRBs above the ankle. Since
hypernovae/under-energetic GRBs belong to the same class as normal
GRBs, one should also expect a natural flux matching  at the ankle
from these two components.}

We have shown that semi-relativistic hypernovae can accelerate CRs
to energies $\lesssim 10^{19}$ eV, and provide the right flux
density { between the second knee at $6\times10^{17}$ eV and
$10^{19}$ eV}. The assumed ejecta velocity distribution profile is
consistent with current supernova-GRB observations. Confirmation
of its theoretical plausibility would require further detailed
numerical investigations of hypernova explosions and shock
propagation through the progenitor envelope.

Recently, there has been evidence indicating that both normal GRBs
and sub-energetic GRBs associated with hypernovae are
preferentially found in low-metallicity galaxies
\cite{low-metallicity}. This would imply that hypernova rates in
normal metallicity galaxies such as our Milky Way may be low, so
that the Galactic contribution to the highest energy CRs would be
unimportant \footnote{A low metallicity host galaxy requirement
for hypernovae is not yet firmly established; without this
requirement, galactic hypernovae might also be  potential high
energy CR sources \cite{Waxman07}, if the galactic rate and the CR
containment time are sufficiently large.}. { Most of the flux
would thus be expected to originate in distant, low-metallicity
galaxies and the distribution of these cosmic rays should be
isotropic. Since the propagation of the cosmic rays in the energy
range $6\times 10^{17}-10^{19}$ eV may be significantly deflected
by the magnetic field in our Galaxy, and possibly by the
intergalactic magnetic field as well, there would be no expected
correlation between the arrival direction of cosmic rays in this
energy range and the hypernova host galaxies. }

We would like to thank E. Waxman, C. D. Dermer, T. K. Gaisser and
P. Sommers for fruitful discussions. This work is supported in
part by NASA NAG5-13286, NSF AST 0307376, and the National Natural
Science Foundation of China under grants 10403002 and 10221001,
and the Foundation for the Authors of National Excellent Doctoral
Dissertations of China (for X.Y.W.).

\end{document}